\documentstyle[amssymb,preprint,aps]{revtex}
\begin{document}
\title{Numerical study of multilayer adsorption on fractal surfaces}
\author{Hao Qi$^1$, Jian Ma$^2$and Po-zen Wong$^1$}
\address{$^1$Department of Physics, University of Massachusetts, Amherst, MA 01003\\
$^2$Department of Physics, Amherst College, Amherst, MA 01002}
\date{October 6, 2000}
\maketitle

\begin{abstract}
We report a numerical study of van der Waals adsoprtion and capillary
condensation effects on self-similar fractal surfaces. An assembly of
uncoupled spherical pores with a power-law distributin of radii is used to
model fractal surfaces with adjustable dimensions. We find that the commonly
used fractal Frankel-Halsey-Hill equation systematically fails to give the
correct dimension due to crossover effects, consistent with the findings of
recent experiments. The effects of pore coupling and curvature dependent
surface tension were also studied.
\end{abstract}

\draft

%TCIMACRO{
%\TeXButton{narrowtext}{\narrowtext%
%}}%
%BeginExpansion
\narrowtext%
%
%EndExpansion

The adsorption of inert molecules on self-similar fractal surfaces has been
a subject of much interest and debate for more than a decade \cite{Reviews}.
It is generally believed that the adsoption isotherm beyond the monolayer
regime follows a scaling form known as the fractal Frankel-Halsey-Hill (FHH)
equation 
\begin{equation}
N=N_{o}[\ln (P_{o}/P)]^{-\zeta }\;,  \label{eq1}
\end{equation}
where $N$ is the number of molecules adsorbed, $P$ is the vapor pressure
above the fractal surface and $P_{o}$ is the saturation vapor pressure at
the temperature of interest. The prefactor $N_{o}$ depends on system
specific parameters such as the linear size of the surface $L$, the density
of the film $\rho $, etc. The central issue has to do with how the exponent $%
\zeta $ depends on the fractal dimension $D$ of the surface. Some authors
have suggested that \cite{PfeiferPRL} 
\begin{equation}
\zeta =(3-D)/3  \label{eq2}
\end{equation}
while others have aruged that \cite{3-D} 
\begin{equation}
\zeta =3-D\;.  \label{eq3}
\end{equation}
The difference has to do with what one believes to be the correct length
scale $\ell $ for measuring the volume of the adsorbed film. If the van der
Waals adsorption potential is dominant, the average film thickness $t$ on a
planar surface is expected to be given by 
\begin{equation}
\alpha a^{3}/t^{3}=k_{B}T\ln (P_{o}/P)\;,  \label{eq4}
\end{equation}
where $a^{3}$ represents the volume occupied by a liquid molecule and $%
\alpha $ is a dimensionless constant. If the surface tension $\gamma $
between the film and the vapor is dominant, the well known Kelvin equation
sets a minimum radius curvature $r$ for the liquid-vapor interface, given by 
\cite{Adamson,Sing} 
\begin{equation}
2\gamma a^{3}/r=k_{B}T\ln (P_{o}/P)\;.  \label{eq5}
\end{equation}
In both cases, $k_{B}T\ln (P_{o}/P)$ represents the chemical potential
difference $(\mu _{o}-\mu )$ between the bulk liquid and vapor phases at
temperature $T$. Since the volume of the wetting film on self-similar
surfaces is expected to scale as $\ell ^{3-D}$ \cite{deGennes}, using either 
$t$ or $r$ as the length scale $\ell $ leads to the two different
predictions for $\zeta $ above. However, Kardar and Indekeu pointed out that
self-affine surface roughness would yet lead to a different prediction and a
correct theory should include both the substrate potential and the surface
tension \cite{Kardar}. In this paper, we report a computer simulation that
is aimed at clarifying the applicability of the theory to real experiments.

Before describing our numerical study, it is necessary to review the status
of the experiments briefly. We note that Eqs. (\ref{eq1})-(\ref{eq3}) have
been widely used to analyze adsorption isotherm data for a long time without
a great deal of success. In many cases, Eq. (\ref{eq1}) fit the data over
less than one decade of length scales \cite{PfeiferPRL,Blacher}. In some
cases, using Eq. (\ref{eq2}) to interpret the exponent gave surface fractal
dimension $D<2$, which is unphysical \cite{Ismail}. While there have been
attempts to compare different methods of analyzing the isotherm data \cite
{Darmstadt}, verification of the results using other techniques such as
small-angle scattering were rare \cite{Sinha,Sahouli}. Thus the
applicability of Eq. (\ref{eq1}) to real systems has remained unclear. To
address this issue, we recently carried out a nitrogen adsorption study of
three shale samples with fractal pore surfaces \cite{Ma}. $D$ for each
sample was independently determined by small-angle neutron scattering (SANS) 
\cite{Wong}. We found that using Eqs. (\ref{eq1})-(\ref{eq3}) to analyze the
adsoprtion data consistently gave lower values for $D$ and there were also
noticeable systematic errors in the fits. Fig. 1 replots the data from one
sample to illustrate the problem. The adsorbed amount is expressed in terms
of the the ratio of the film volume $V_{ad}$ to the total pore volume $%
V_{full}$. The length scale is expressed in terms of the Kelvin radius $r$
measured in units of $a$ \ (0.4 nm for N$_{2}$ molecules) and it is
calculated using Eq. (\ref{eq5}) for a range of pressure $P$ relative to the
saturation value $P_{o}$ for liquid nitrogen at 65 K, the temperature of the
experiment. The fit was made over a two-decade range (0.4 - 50 nm) that
coincides with the SANS data. Fig. 1a shows that using a linear fit on
log-log scales and interpreting the slope ($=0.46$) according to Eq. (\ref
{eq3}) give $D=2.54$, a result that is significantly less than the value $%
2.83$ obtained by SANS. Using Eq. (\ref{eq2}) for the slope would give an
unphysical value of $1.62$.\ The horizontal dotted line at $%
V_{ad}/V_{full}\thickapprox 0.1$ represents the monolayer capacity as
determined by the standard BET analysis \cite{Sing} and the vertical dotted
line at $r/a\thickapprox 3$ represents the onset of hystereses between
adsorption and desorption (not shown). The latter is a signature of
capaillary condensation in small pores due to surface tension \cite{Barrer}.
There is clearly a shoulder above the fitted line near this point where the
adsorbed volume is still less than two monolayers. At higher pressure or
long length scales where capillary condensation is expected to be more
dominant, the data shows a slight upward curvature. Fig. 1b shows that
fitting only the data above the shoulder to Eq. (\ref{eq1}) with an additive
background gives $D=2.37$, if Eq. (\ref{eq3}) is used, which makes the
disagreement with SANS worse.\ The same systematic deviations were seen in
all three samples with different fractal dimensions \cite{Ma}. The most
likely explanation is that Eqs. (\ref{eq1}) - (\ref{eq3}) represent
asymtotic behavior that may not be observable in real systems. The crossover
between van der Waals adsorption and capillary condensation may not permit
such a simple analysis. The best way to test this hypothesis is to carry out
a computer simulation in a simple model with a known fractal dimension.

The model we used to represent a fractal surface is an assembly of
independent spherical pores with a power law distribution in the radii $R$ .
If the number density follows $g(R)\propto R^{-(D+1)}$, then the surface
area measured on a length scale $\ell $ is $S_{\ell }\propto \int_{\ell
}R^{2}g(R)$d$R$ $\propto \ell ^{2-D}$, as expected of a fractal surface.
Hence $D$ can be easily adjusted without any complication in the
calculation. For each pore radius $R$, how the film thickness $t$ grows with
increasing pressure $P$ up to the point of capillary condensation can be
calculated numerically by combining Eqs. (\ref{eq4}) and (\ref{eq5}) to
account for both the substrate potential and the surface tension. The
resulting equation in dimensionless form is 
\begin{equation}
\ln (P_{o}/P)=\frac{\mu _{o}-\mu }{k_{B}T}=B\left[ \frac{a^{3}R^{3}}{%
t^{3}(2R-t)^{3}}+\frac{Ca}{R-t}\right]  \label{eq6}
\end{equation}
where $B=8\alpha /k_{B}T$ and $C=\gamma a^{2}/4\alpha $. The first term on
the right comes from the van der Waals potential inside a spherical pore of
radius $R$ at a normal distance $t$ from the surface and $B$ is a measure of
the adsorption energy relative to the thermal energy. The second term is due
to the surface tension of a meniscus of radius $R-t$\ and $C$ is a measure
of the relative strength of surface tension to the substrate potential. In
real systems, $B$ and $C$ should be of order unity or greater when
adsorption and capillary condensation occur.

The solution of Eq. (\ref{eq6}) can be best understood using the graphical
approach of Cohen {\it et al }\cite{Cohen}. Fig.2 is a plot of the equation
with $B=C=1$ and $R/a=10$. The solid line shows how the right-hand side of
Eq. (\ref{eq6}) varies with the film thickness $t/a.$ The contributions from
the two separate terms are indicated by the dashed lines. We note that while
the van der Waals term falls off with increasing $t$, the surface tension
term rises with $t$. Hence their sum has a minimum at the point $%
(t_{c},P_{c})$. For any nonzero pressure $P$, a horizontal (dotted) line can
be drawn through the point $\ln (P_{o}/P)$ on the vertical axis. For low
pressures, the line intersects the solid curve at two points that are the
solutions of Eq. (\ref{eq6}), but the physical film thickness is given only
by the smaller of the two $t$ values.\ With increasing pressure, the
horizontal line is shifted downward and $t$ increases. At the critical
pressure $P_{c}$, the line is tangent to the solid curve and it gives the
maximum film thickness $t_{c}$, which is the threshold of condensation. At
pressure above $P_{c}$, Eq. (\ref{eq6}) has no real roots for $t$, i.e., the
pore is filled. We can see from Fig. 2 and Eq. (\ref{eq6}) that increasing $%
B\ $\ would shift both of the dashed lines upward, thereby lowering $P_{c}$
without affecting $t_{c}$. Increasing $C$ would shift only the dashed line
on the right, thereby lowering both $P_{c}$ and $t_{c}$. In our study, we
varied both $B$ and $C$ over the range of $0.1-10$. Each choice of $(B,C)$
gives a different isotherm. We chose three different $D$ values ($2.5$, $2.7$
and $2.85$) to roughly match our earlier experiment \cite{Ma}.

To simulate an adsorption isotherm, we used $200$ pore radii distributed
uniformly on a logarithmic scale over three decades: $2\leq R/a\leq $ $2000$%
. For each $R$, Eq. (\ref{eq6}) can be solved to find the condensation
pressure $P_{c}$. In practice, we found for each pressure $P$ the critical
radius $R_{c}$. All the pores with $R<R_{c}$ are filled, and for those with $%
R>$ $R_{c}$, the film thickness $t$ was calculated. The total adsorbed
amount was obtained by summing the contributons over the power-law
distribution $g(R).$ Fifty pressure steps were used to obtain the entire
isotherm and the pressure range was chosen such that the Kelvin radius $r$
in Eq. (\ref{eq5}) spanned four decades ($0.1a$ to $1000a$), much wider than
any real experiment. In Fig. 3a, the open circles depict the simulated
isotherm with $B=C=1$ for a sample with $D=2.85$, a value that matches
closely to the real sample in Fig.1. The simulation data exhibit a shoulder
just above $r/a\thickapprox 1$ in a manner similar to the real data in Fig.
1, but here we can see explicitly that it caused by the onset of capillary
condensation in the smallest pore: above this threshold, the increase of
pressure (or $r$) causes a rapid rise in condensed volumes (solid triangles)
and a corresponding decrease of film volume (crosses).\ That the adsorption
is separated into these two components explains why using either Eq. (\ref
{eq2}) or Eq.(\ref{eq3}) alone to interpret the data over the entire range
would result in systematic errors. At best, one should expect Eq. (2) to
apply below the condensation threshold and Eq. (3) above it.

Figure 3b shows how the local slope of the simulated isotherm varies in a
log-log plot. In the limit of small $r$ (Region I) where there is no
condensation, the slope of $0.35\thickapprox 1/3$ corresponds to $%
D\thickapprox 2$ according to Eq. (\ref{eq2}). This proved to be a robust
result for all the cases we examined. It implies that the film thickness $t$
given by Eq. (\ref{eq4}) is not a length scale that could be used to probe
fractal features, because $t$ can be used only when the system is free of
capillary condensation, but that guarantees it to be below the lower cut-off
of the fractal features (the smallest pore size). As a\ result, the only
possiblilty is to use the Kelvin radius $r$ and analyze the data in the
limit of large $r$, because the volume of the adsorbed film is a smaller
percentage of the pore volume in the larger pores. In our simulation, we
found that this approach works only for cases with large values of $C$ and
only near the upper limit of the range of $r$ we used. In other cases, we
still found fractal dimensions clearly below the true value. For example,
the data in Fig. 3b fit a slope of $0.25$ in the large $r$ limit (Region
III). It corresponds to $D=2.75$ according to Eq. (\ref{eq3}), which is
appreciably less than the input value of $2.85$ used for the simulation. In
the middle region (Region II) where $r$ is comparable to the real data in
Fig.1, the slope is higher ($0.36$) and it results in an even lower observed
dimension ($D=2.64$). This trend of disagreement is consistent with what we
found between our SANS and adsorption experiments \cite{Ma}. Similar
behavior was observed in our simulated samples with $D=2.7$ and $D=2.5$.
There is always an anomaly at the onset of capillary condensation, albeit
the appearance varies somewhat. The reason is that the slope in Region I is
always close to $1/3$ while that in Region III varies with $D$. So how the
data join together in the middle is not universal. For the $D=2.5$ case, the
slope in Region III tends to $0.5$, which is much higher than that in Region
I. Hence the shoulder structure is masked by the intrinsic increase of the
slope from $1/3$ to $1/2$, making the anomaly appear more like a kink. For $%
D=2.7$, although the slope in Region III tends to $0.3$, a value nearly
identical to that in Region I, the shoulder appearance remains pronounced.

Our model was extended in an {\em ad hoc} manner to include some effects
that occur in real systems. For example, the density of a liquid film is
typically higher near the substrate and liquid/vapor surface tension is
expected to be stronger for concave surfaces with small radius of curvature.
Both of these effects would increase the second term in Eq. (\ref{eq6})
relative to the first term. We investigated these effects qualitatively by
letting $C$ be a function $C(R)=C_{1}+C_{2}\exp (-R/\lambda )$. From the
data shown in Ref. \cite{Sing}, we estimated that $\lambda \thickapprox 5.4a$%
. In addition, for any real fractal surface, features of different sizes are
not completely independent. Condensation in small pores can trigger the same
to occur in neighboring large pores. We modelled such pore coupling effects
by letting pores with radius $R>R_{c}\ $to condense with a probability $%
K(R_{c}/R)^{q}$, i.e., less likely for larger pores. Fig. 3c shows that with 
$C_{1}=C_{2}=1,$ $K=0.3$ and $q=4$, the isotherm in Fig. 3b acquires a more
pronounced shoulder. The curvature in Region II also changes from concave
downward to concave upward. A power law fit gives a slightly lower dimension
($D=2.60$). These results are in qualitative agreement with the real data in
Fig. 1. They can be explained by the fact that the increased surface tension
makes condensation in small pores to occur at a lower pressure and pore
coupling makes some of the larger pores to follow. The net result is to
enhance the anomaly at the onset of capillary condensation.

Despite the fact that our model is far from being a realistic representation
of any real system, it clarifies a long-standing ambiguity associated with
the interpretation of adsorption isotherm on fractal surfaces. By varying
the parameters $B$, $C$ and $D$, we are able to see explicitly how the
crossover between substrate driven adsorption and surface tension driven
condensation prevents one from using the FHH equation to obtain the correct
fractal dimension under common experimental conditions. We are also able to
produce simulated isotherms that resemble the real data and understand the
cause of their essential features. We hope that these understandings, though
only qualitative, will serve as useful guides for interpreting adsorption
isotherms with the FHH equation.

We thank J. Machta and R. Guyer for helpful discussions. This work was
supported by The Petroleum Research Fund administered by the American
Chemical Society under Grant Nos. 32191-AC2-SF98 and 33549-B9, and NSF Grant
No.\ CTS-9803387.

\begin{figure}[tbp]
\caption{Nitrogen adsoption isotherm of a shale sample with fractal pore
surface. The fractal dimension $D$ obtained by SANS is 2.83 but analyses
using Eqs. (\ref{eq1}) and (\ref{eq3}) give much lower values. (a) A linear
fit on a log-log scale over two decades of length scales gives $D=2.54$. (b)
A power law fit of the data in the capillary condensation regime gives $%
D=2.37$.}
\label{fig1}
\end{figure}

\begin{figure}[tbp]
\caption{A graphical representation of the solution of Eq. (\ref{eq6}). The
two dashed lines represent the two terms on the right-hand side of the
equation and the solid curve is their sum. The minimum at ($t_{c},P_{c}$)
corresponds to the threshold of capillary condensation in a pore of radius $%
R $ for the given parameters $B$ and $C$. See text for details.}
\label{fig2}
\end{figure}

\begin{figure}[tbp]
\caption{Simulated isotherms for a sample with $D=2.85$ and $B=C=1\,$\ (a)
The onset of capillary condensation causes the appearance of a shoulder near 
$r/a=1$. (b) The variation in the local slope of the isotherm is due to the
crossover from the van der Waals effects (Region I) to surface tension
effects (Region III). The slope in the two limits should be given by Eqs.
(2) and (3) respectively. Most experiments fall in between (Region II). The
correct fractal dimension is not observed in any of the three regions in
this example. (c)\ By letting the constant $C$ be $R$-dependent and pores
with radius $R>R_{c}$ to condense with a probability $K(R_{c}/R)^{q}$, the
shoulder in the isotherm is accentuated.}
\label{fig3}
\end{figure}

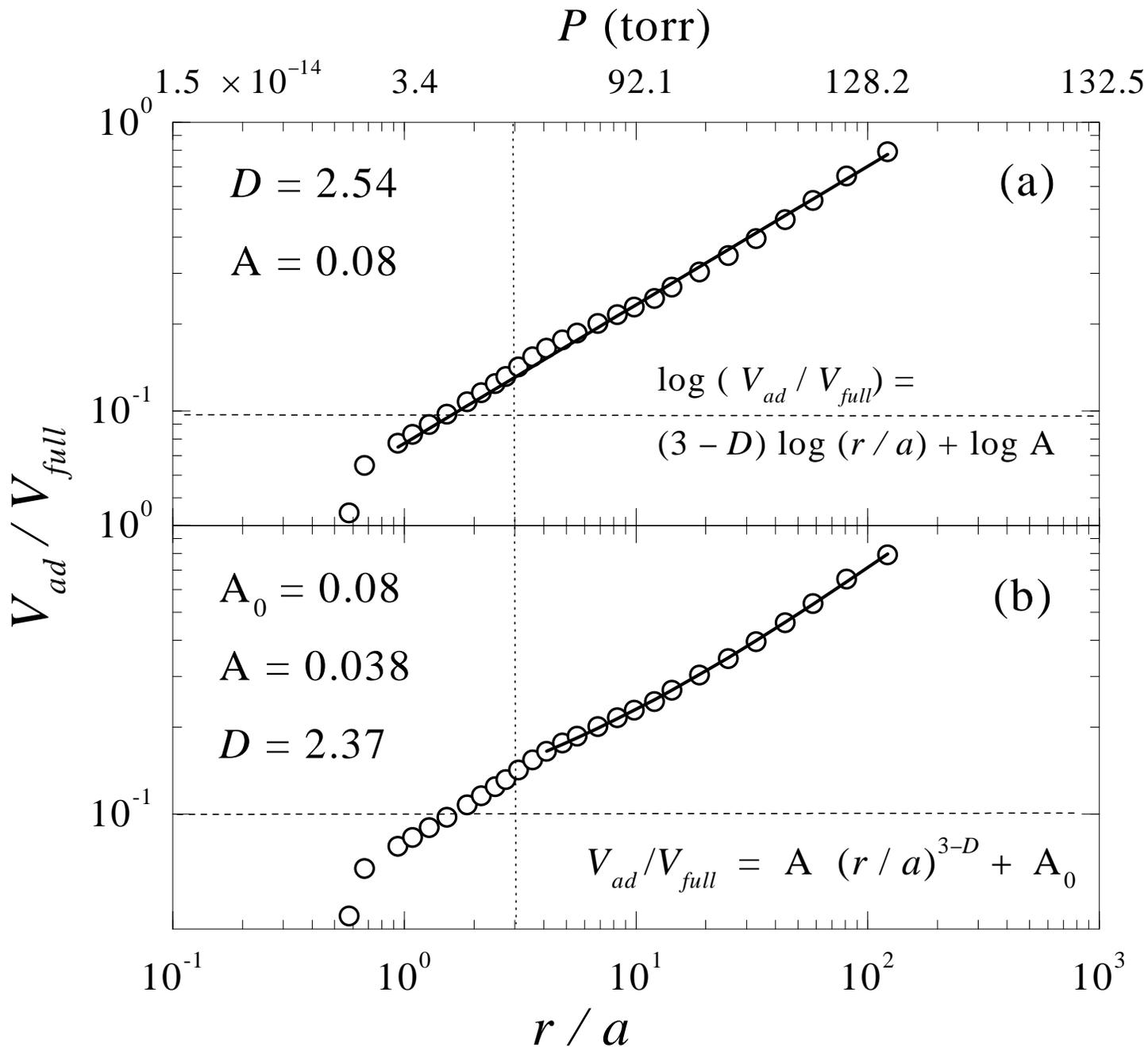
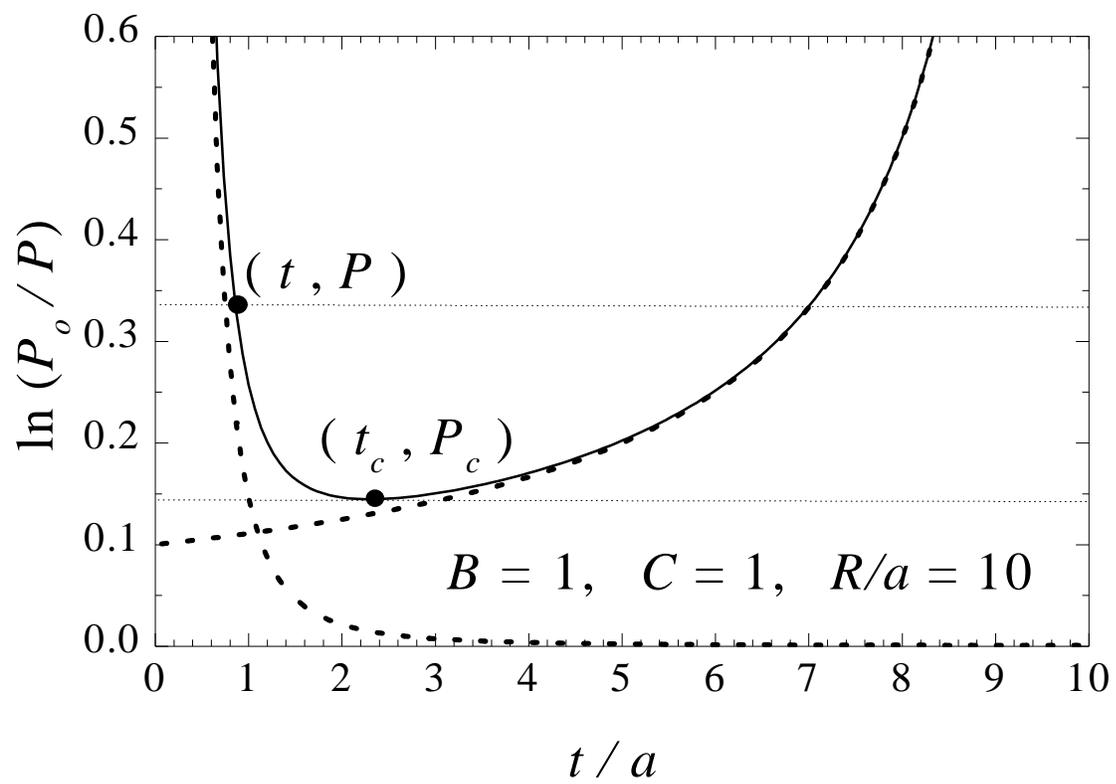
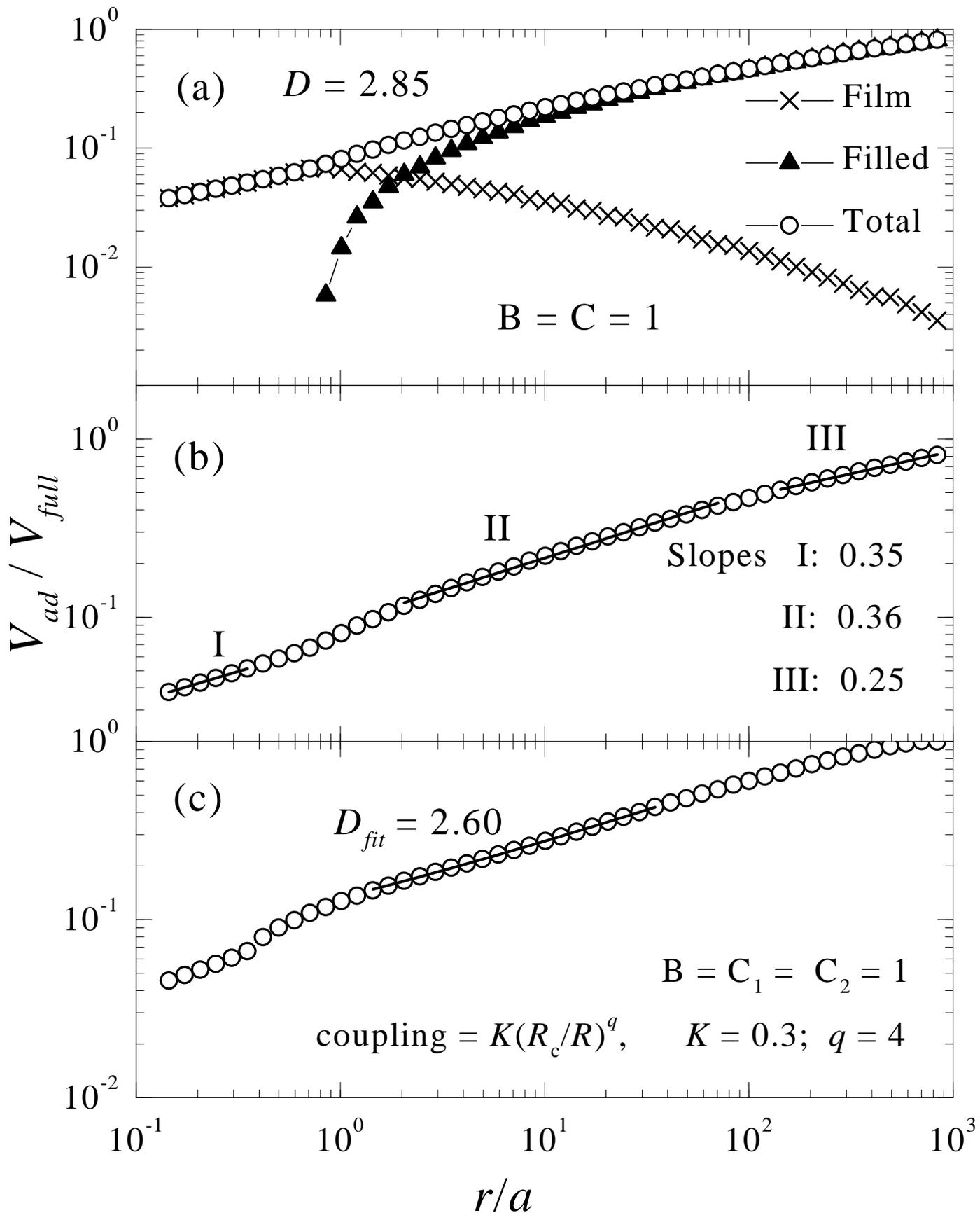
\end{document}